\documentstyle[proc]{rspublic}
\include{psfig}
\def\bi{\begin{itemize}}
\def\ei{\end{itemize}}
\def\bq{\begin{quotation}}
\def\eq{\end{quotation}}

\def\thedemobiblio#1{\smallskip\par
 \list{}{\labelwidth 0pt \leftmargin 1em \itemindent -1em \itemsep 1pt}
 \small \parindent 0pt
 \parskip 1.5pt plus .1pt\relax
 \def\newblock{\hskip .11em plus .33em minus .07em}
 \sloppy\clubpenalty4000\widowpenalty4000
 \sfcode`\.=1000\relax}

\begin{document}

\title[The AAT 2dF QSO redshift survey]{
QSO clustering and the AAT 2dF QSO redshift survey}

\author[B.J. Boyle et al.]{B.J.Boyle$^1$, S.M.Croom$^2$, R.J.Smith$^3$,
T.Shanks$^2$, L.Miller$^4$, N.Loaring$^4$}

\affiliation{
1. Anglo-Australian Observatory, 
PO Box 296, Epping, NSW 2121, Australia\\
2. University of Durham, South Road, Durham DH1 3LE, UK\\
3. Institute of Astronomy, Madingley Road, Cambridge CB3 0HA, UK\\
4. University of Oxford, 1 Keble Road, Oxford OX1, UK
}

\maketitle

\section{Introduction}

The study of QSO clustering from large homogeneous surveys is likely
to provide unique information on the nature of large-scale structure
(LSS) in the Universe.  Even with relatively modest aperture
telescopes, QSOs are readily detectable from the present cosmic epoch
to a time when the Universe was less than 25\% of its present age.
This approximately eight billion year time interval provides a unique
baseline over which to study the evolution of structure in the
Universe with cosmic time.  QSO clustering also yields a direct
measure of clustering on comoving scales 
$\gg 100\,$h$^{-1}$Mpc\footnote{\hangindent=1.37pc\hangafter=1
Throughout this paper we use comoving distances, $r_{\rm comoving} 
= (1+z)r_{\rm proper}$, and define the Hubble constant
as $H_0=100$h$\,$km$\,$ s$^{-1}$Mpc$^{-1}$}, 
providing key information
(see figure 1) on LSS at scales between those provided by the COBE
results ($>1000\,$h$^{-1}$Mpc, Bennett {\it et al.} 1994) and galaxy
redshift surveys ($<100\,$h$^{-1}$Mpc, Colless {\it et al.}, this
volume)

\begin{figure}
\centering \centerline{\psfig{file=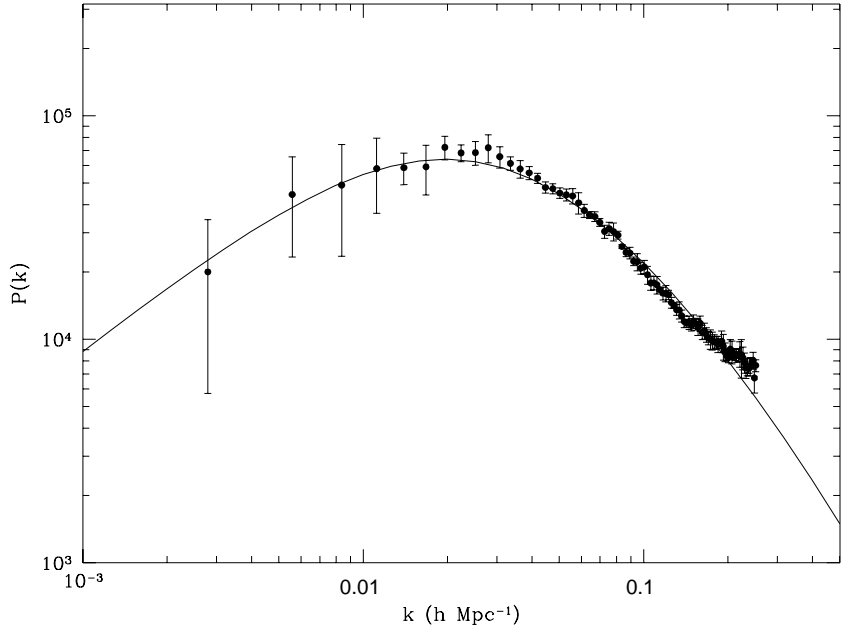,width=12cm}}
\caption{The QSO power spectrum, P(k), based on a Zel'dovich
approximation in a (2250h$^{-1}$Mpc)$^3$ box containing 192$^3$
particles with $z_{\rm max}=2.2$.  A $\tau$CDM model with $\Omega_0=1$,
$\Gamma=0.2$ $\sigma_8=0.67$ and QSO bias $b_{\rm Q} =1.8$ at $z=0$ is
assumed.  Errors on P(k) are based on the rms scatter from five
simulations.  Note the 2dF QSO redshift survey comprises two such 
slices, effectively reducing errors by 1.4. Simulation taken from
Croom et al.\ (1998, in preparation).}
\end{figure}

To date, QSO clustering has remained largely unexploited for LSS
studies.  This has largely been due to the lack of appropriate QSO
catalogues.  Although over 10000 QSOs with measured redshifts are now
known (V\'eron \& V\'eron-Cetty 1997, see figure 2) less than $20\,$\%
of these QSOs form part of homogeneous catalogues suitable for
detailed clustering analysis.  The goal of the 2dF QSO redshift
survey, currently underway at the Anglo-Australian Telescope (AAT), is
to provide a new homogeneous catalogue of almost 30000 QSOs with which
to carry out such an analysis.  The survey is almost two orders of
magnitude larger than the previous largest homogeneous survey to this
depth (Boyle {\it et al.}\ 1990).

In \sect2 we review the existing results on QSO clustering and we
describe the creation of the 2dF QSO redshift survey input catalogue
in \sect3. We report on the initial results from the survey in
\sect4 and discuss the future prospects for this and other
QSO surveys in \sect5.

\begin{figure}
\centering \centerline{\psfig{file=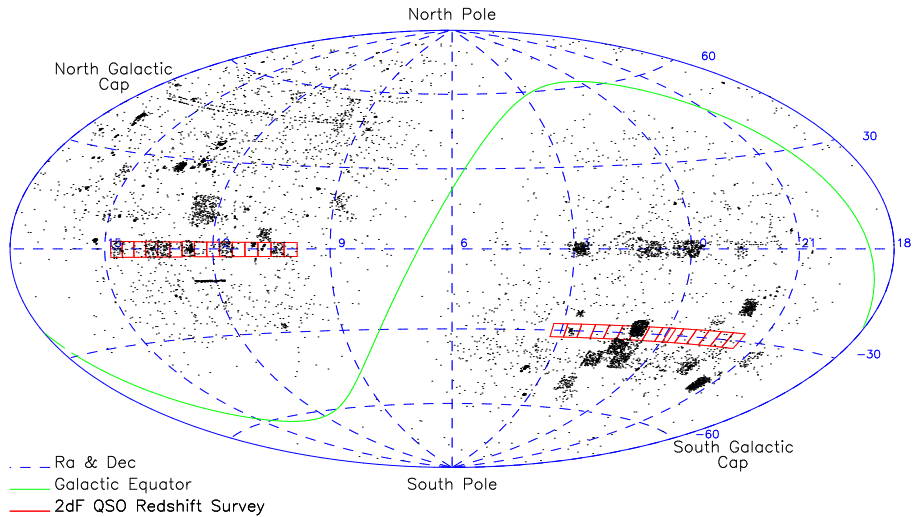,angle=-90,width=12cm}}
\caption{The all-sky distribution of the $\sim 10000$ known QSOs from
the catalogue of V\'eron \& V\'eron-Cetty (1997).  The thirty UKST fields
of the 2dF QSO redshift survey are also shown in their two declination 
strips.}  
\end{figure}

\section{QSO Clustering}

\subsection{QSO-QSO clustering}

First attempts to detect QSO clustering from samples of QSOs yielded
no significant detections of QSO clustering on any scale up to
1500h$^{-1}$Mpc (see e.g. Osmer 1981). Nevertheless, a number of
studies (Osmer 1981, Webster 1982) did comment on the possibly genuine
nature of specific clusters of QSOs at intermediate scales $\sim
50-100$h$^{-1}$Mpc.  The first positive detection of clustering
amongst QSOs at small scales ($<10$h$^{-1}$Mpc) was made by Shaver
(1984) using an earlier version of the heterogeneous V\'eron \&
V\'eron-Cetty (1997) catalogue, which then comprised 2000 QSOs.  This
was followed by similar results based on more homogeneous QSO
catalogues, comprising a few hundred QSOs (Shanks, Boyle \& Peterson
1988, Iovino \& Shaver 1988).  The strength of QSO clustering in these
studies was found to be similar to that of galaxies at the present
day, with the two-point QSO auto-correlation function, $\xi_{\rm
Q}(r)$, fitted by a $-1.8$ power law $\xi_{\rm Q}(r) = (r/r_0)^{-1.8}$
with a scale length $r_0 \sim 6\,$h$^{-1}$Mpc.  None of these early
QSO clustering studies found any significant QSO clustering on scales
$>10$h$^{-1}$Mpc.

Suitable QSO sample sizes have only increased slowly over the past ten
years, and latest analyses of QSO clustering (see e.g. Shanks \& Croom
1996) still rely on a compilation of a few homogeneous catalogues
(Durham/AAT: Boyle {\it et al.} 1990, CFHT: Crampton {\it et al.}
1989, LBQS: Hewett {\it et al.} 1995) comprising less than 2000 QSOs
in total.
\begin{figure}
\centering \centerline{\psfig{file=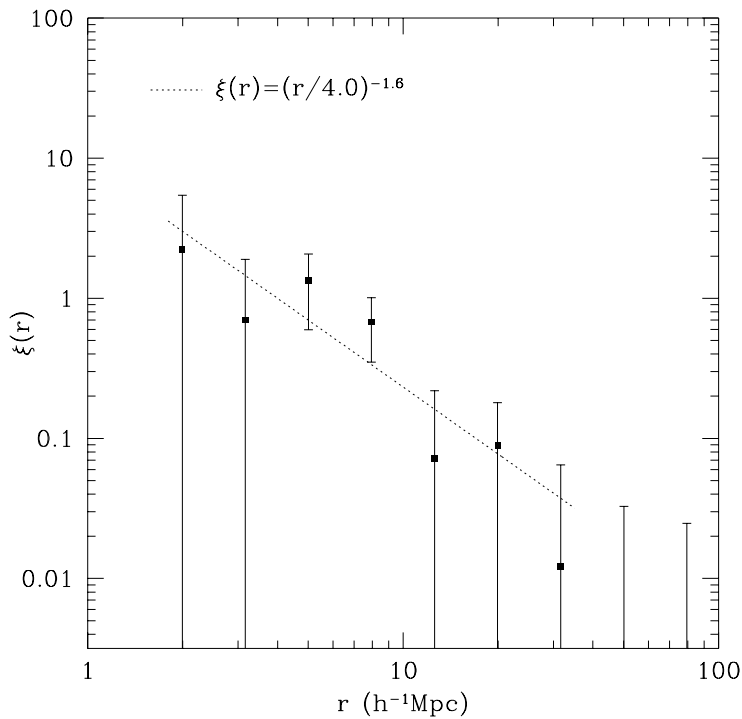,width=10cm}}
\caption{The QSO auto-correlation function $\xi_{\rm Q}(r)$ for the
redshift range $0.3<z<2.2$ derived by Croom \& Shanks (1996) from the
Durham/AAT (Boyle {\it et al.} 1990), CFHT (Crampton {\it et al.}
1989) and LBQS (Hewett {\it et al.} 1995) surveys.  The total
sample comprises $\sim1500$ QSOs.}
\end{figure}

Using the Durham/AAT, CFHT and LBQS catalogues, Croom \& Shanks (1996)
demonstrated that QSO auto-correlation function at scales less than
$30$h$^{-1}$Mpc was best-fit by $\xi_{\rm Q}(r) =
(r/r_0)^{-1.59\pm0.46}$, with $r_0 = 4.0 \pm 1.1\,$h$^{-1}$Mpc (figure
3).  Croom \& Shanks (1996) also found no evidence for any evolution
in the strength of clustering with redshift. By investigating the
predications of a standard biasing model, they were able to
demonstrate that the low values of $r_0$ obtained for $z<0.2$ Seyfert
galaxies (Boyle \& Mo 1993) and IRAS-selected AGN (Georgantopoulos \&
Shanks 1994) favoured either a low $\Omega_0$ universe, $\Omega_0 <
0.1$, with low values for the galaxy bias parameter ($b_{\rm g} \sim
1$) or an $\Omega_0 = 1$ universe with a higher bias parameter
($b_{\rm g} \sim 2$). On the basis of this model, Croom \& Shanks were
also able to rule out a universe with $\Omega_0 = 1$, $b_{\rm g} = 1$ at
the $\sim 3\sigma$ level.

These results on the evolution of the QSO correlation function are,
however, contested by other authors. In their earlier analysis, Iovino
\& Shaver (1988) found that the correlation length decreased towards
high redshift, while more recently La Franca {\it et al.}  (1997) claim
to see an increase in the QSO correlation length with redshift, with
$r_0 =4.2$h$^{-1}$Mpc for $z<1.4$ QSOs and $r_0 = 9.1$h$^{-1}$Mpc for
$z>1.4$ QSOs.

The lack of agreement between these results serves to underline the
poor statistics on which the detection of QSO clustering at small
scales is currently based.  There are less than 40 pairs of QSOs with
comoving separations $r<10$h$^{-1}$Mpc identified in such clustering
studies, even when averaged over the full range in redshift 
$0.3<z<3$. 

At intermediate scales, $30$h$^{-1}$Mpc$ < r < 100$h$^{-1}$Mpc, there
have been a number of claims for marginal ($2\sigma$) detections of
features in the QSO auto-correlation function.  However, none of these
has stood up to critical examination with larger data-sets (see Croom
\& Shanks 1996).  At these scales, the growth of LSS largely follows
linear theory and thus features in $\xi_{\rm Q}(r)$ at these comoving
separations should not change comoving scale as a function of
redshift.  If detected, such features would therefore provide a useful
cosmological `standard rod' with which to measure the deceleration
parameter $q_0$ or the cosmological constant $\lambda_0$.

Croom and Shanks (1996) also compared $\xi_{\rm Q}(r)$ in the
$8-50\,$h$^{-1}$Mpc region with an extrapolation of the CDM power
spectrum normalised by the COBE data.  This directly compares the
fluctuations in the QSO distribution with those in the mass
distribution and is thus a potentially powerful way to measure
directly the QSO bias parameter.  For a standard CDM model with
$\Omega_0 =1$, they obtained a QSO bias of $b_{\rm
Q}(z=0)=1.40^{+0.28}_{-0.43}$.  For a CDM model with a cosmological
constant ($\Omega_0 =0.2$, $\lambda_0=0.8$,
$\Gamma=0.2\sim\Omega_0$h), they obtained $b_{\rm
Q}(z=0)=1.20^{+0.13}_{-0.18}$.

The inconsistency between the small scale results, which favour a high
bias for $\Omega_0=1$, and the COBE comparison, which favour a low
bias for $\Omega_0=1$, could be taken to mean that a low value for
$\Omega_0$ may be more appropriate.  However, the errors are still
large, and it will take surveys such as the 2dF QSO survey to reduce
them to the point where more effective discrimination between these 
models is possible.

Phillips {\it et al.} (1996) and Ballinger {\it et al.} (1997) have
also developed a potentially useful geometric test which may be
employed at these scales to place limits on the value of $\lambda_0$.
The test relies of the fact that, on average, QSO clusters should be
spherically symmetric i.e. with as large an extent along the line of
sight as transverse to it. Since $\lambda_0$ has a different effect on
these two distances, clusters will appear to be spherically symmetric
only when the correct value of $\lambda_0$ is used.  Ballinger {\it et
al.}  (1996) demonstrated that the 2dF QSO redshift survey would indeed
be able to measure this effect, although effects such as
redshift-space distortions serve to weaken the signal and complicate
the analysis.

At the largest scales, all measures of the correlation function give a
signal consistent with zero clustering.  At comoving separations
greater than 100h$^{-1}$Mpc Croom \& Shanks (1996) find $\xi_{\rm Q}
(r) = 0 \pm 0.025$.  Isolated examples of QSO `superclusters' with
dimensions $\sim 100-200$h$^{-1}$Mpc have previously been identified
(Crampton {\it et al.} 1991, Clowes \& Campusano 1993), but it is
unclear how representative they are of the QSO population as a whole.

\subsection{QSO-galaxy clustering}

A limiting factor in the study of LSS with QSOs is the 
unknown bias factor which exists between 
the QSO population and the underlying mass distribution. 
Studies of the galaxy clustering around QSOs provide a useful
constraint on the relationship between the QSO and galaxy bias. 

The study of the galaxy environment around QSOs was first put 
on a sound statistical footing by Stockton (1982) and later 
developed by Yee and his co-workers in a series of papers,
see Ellingson {\it et al.} (1991) and references therein.

The work of Ellingson {\it et al.} (1991) largely focussed on
radio-loud QSOs, and demonstrated that their galaxy environment
undergoes a dramatic increase in richness from low redshifts to
intermediate redshifts, corresponding to an evolution in the galaxy
clustering scale length around QSOs from $r_0 \sim 5$h$^{-1}$Mpc
(consistent with field galaxies) at $z\sim 0.3$ to $r_0 \sim
15$h$^{-1}$Mpc (similar to Abell richness class 1) at $z\sim 0.6$.
Evidence for similar levels of strong galaxy clustering around
radio-loud QSOs at yet higher redshifts ($z\sim 1$) have been found by
Tyson (1986) and Hintzen {\it et al.} (1991).

\begin{figure}
\centering \centerline{\psfig{file=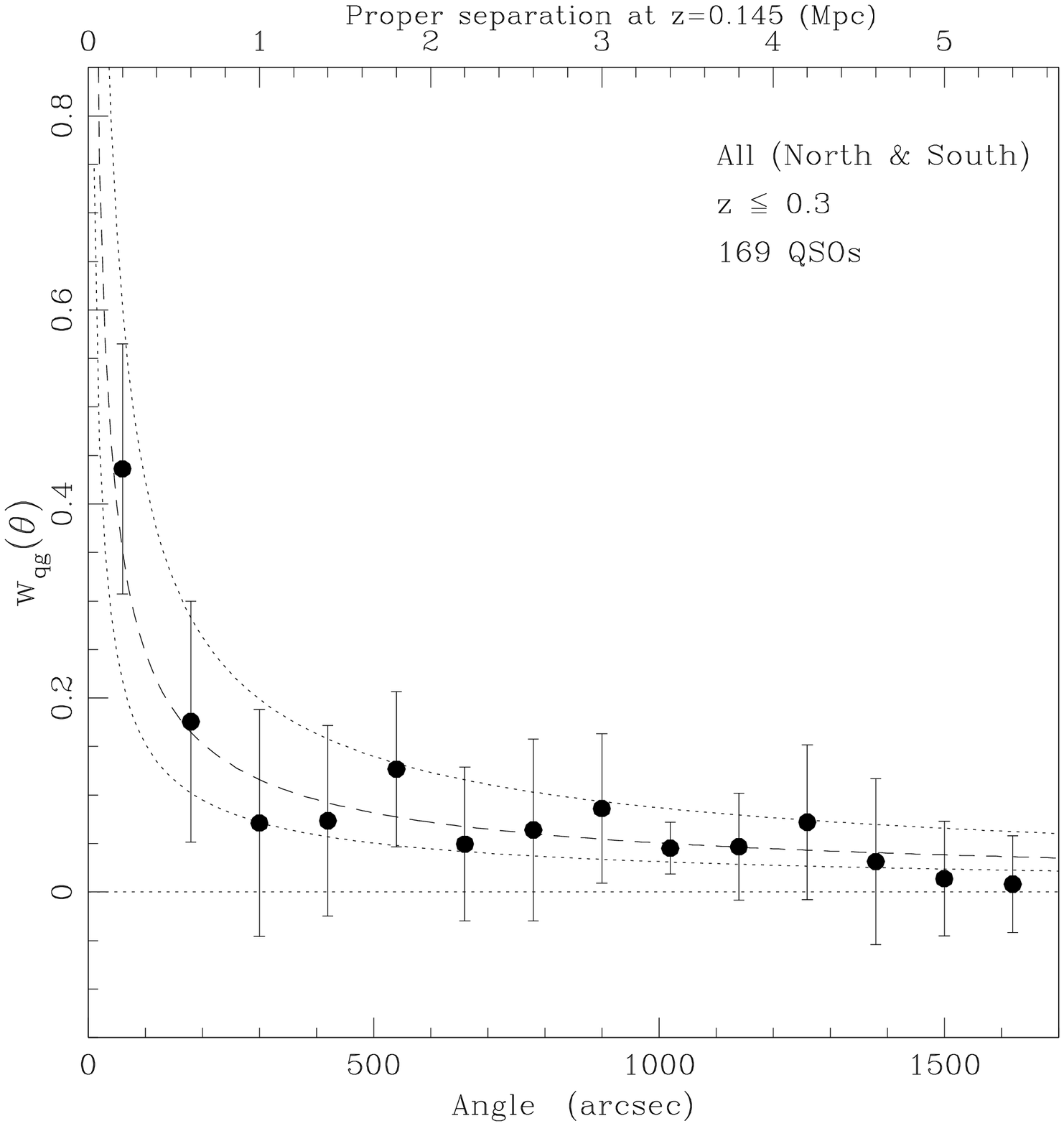,width=9cm,height=8cm}}
\centering \centerline{\psfig{file=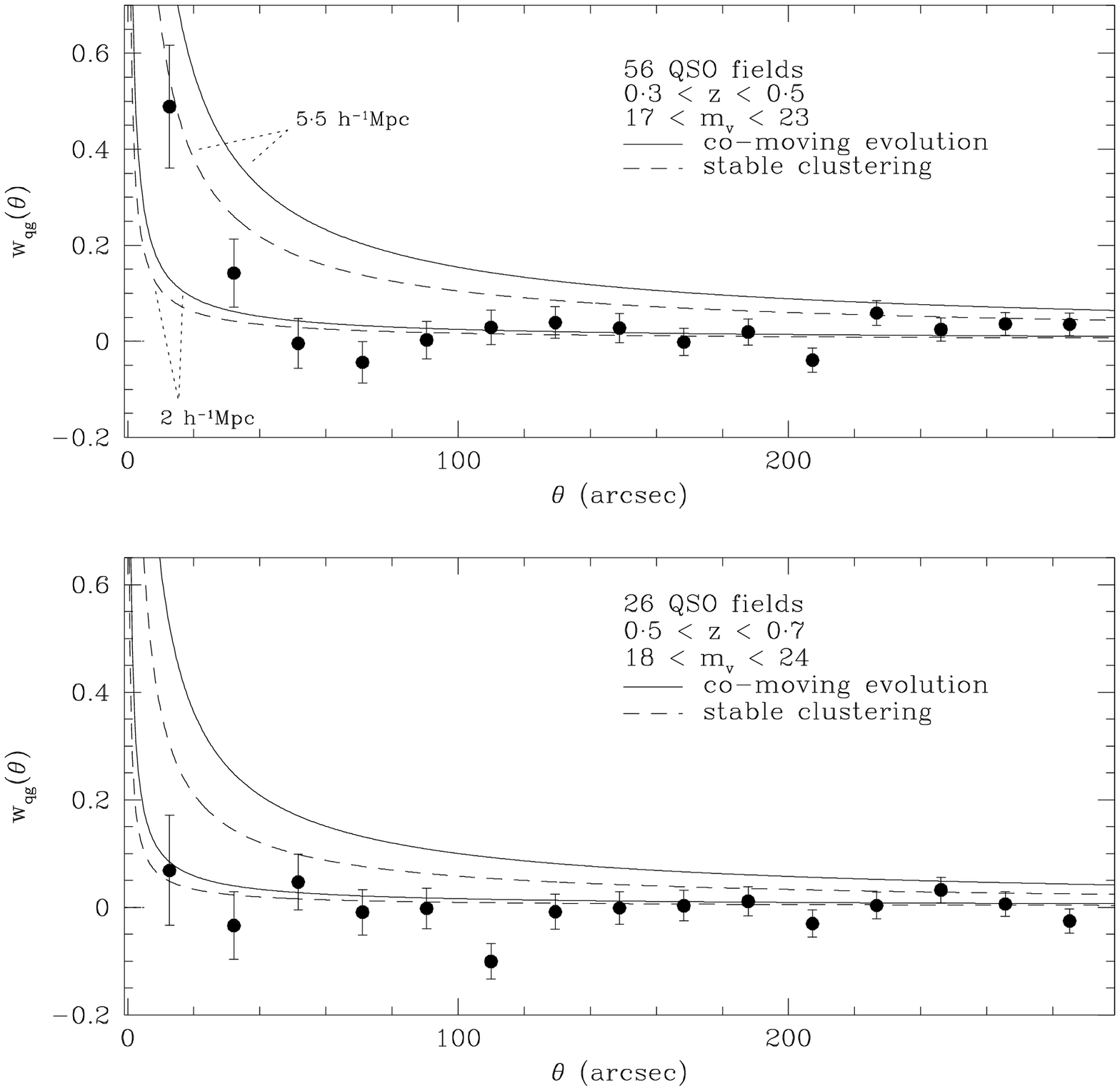,width=10cm}}
\caption{The QSO-galaxy cross-correlation functions, $w_{\rm
qg}(\theta)$, for samples of X-ray selected QSOs at $z<0.3$ (top),
$0.3<z<0.5$ (middle), and $0.5<z<0.7$ (bottom). In the $z<0.3$ sample
the $w_{\rm qg}(\theta)$ is directly compared with the galaxy
auto-correlation function (dashed line).  In the higher redshift
samples predicted cross-correlations functions are shown for
QSO-galaxy scale lengths of $r_0=2$h$^{-1}$Mpc and
$r_0=5.5$h$^{-1}$Mpc under certain assumptions regarding the evolution
of clustering.  See Smith {\it et al.} (1996) and Smith (1998).}
\end{figure}

In contrast, the environments of the vast majority of QSOs which
are radio-quiet appear to 
significantly poorer.  Smith {\it
et al.} (1996) demonstrated that the QSO-galaxy cross-correlation
function for a sample of over 100 X-ray selected radio-quiet AGN at
$z<0.3$ from the Einstein Medium Sensitivity Survey was identical to
the galaxy auto-correlation function over a similar range in redshift.

Smith (1998) has now extended this analysis to higher redshift QSOs in
the same X-ray selected sample, finding evidence for the evolution in
the QSO-galaxy cross-correlation function in the opposite sense to
that witnessed for radio-loud QSOs. The QSO-galaxy scale length
decreases from $r_0=5$h$^{-1}$Mpc at $z<0.3$ to $r_0\sim2$h$^{-1}$Mpc
in the redshift range $0.5<z<0.7$ (figure 4).  The inferred evolution
of the QSO-galaxy cross-correlation function may be expressed in the
form $\xi_{\rm QG}(r) \propto r^{-1.8}(1+z)^{-3+\epsilon}$, with
$\epsilon \sim 0.8$, equivalent to the growth of structure predicted
by linear theory in an $\Omega_0=1$ Universe (Peacock 1997).

This evolution is also similar to the evolution of the galaxy
scale length as inferred indirectly from the galaxy angular
auto-correlation function derived from deep imaging surveys (see
Efstathiou 1995) and measured directly from deep redshift surveys (Le
F\`evre {\it et al.} 1996).

Based on these results we can infer that radio-loud and radio-quiet
QSOs inhabit different galaxy environments, probably indicative of
different levels of bias between the two populations.  This is further
borne out by the QSO correlation length ($r_0 \sim 18$h$^{-1}$Mpc)
inferred from two-dimensional radio-source catalogues (Loan {\it et
al.} 1997).  This level of clustering is more consistent with recent
estimates of the galaxy cluster correlation length ($r_0 \sim
14$h$^{-1}$Mpc, Dalton {\it et al.} 1994).  However, the extent to
which the radio-quiet QSO population is representative of the galaxy
population will probably only be answered by detailed
redshift-dependent QSOs clustering studies in the 2dF QSO redshift
survey.

\section {The 2dF input catalogue}

\subsection{Creation} 

The identification of the QSO candidates for the 2dF redshift survey
was based on broadband $UB_JR$ colour selection from APM measurements
of UK Schmidt (UKST) photographic plates.  The survey area comprises
30 UKST fields, arranged in two $75^{\circ} \times 5^{\circ}$
declination strips centred on $\delta=-30^{\circ}$ and
$\delta=0^{\circ}$ (see figure 5). The $\delta=-30^{\circ}$ strip
extends from $\alpha = 21^{\rm h}40$ to $\alpha = 3^{\rm h}15$ and the
equatorial strip from $\alpha = 9^{\rm h}50$ to $\alpha = 14^{\rm
h}50$. The total survey area is $740\,$deg$^2$, when allowance is
made for regions of sky excised around bright stars, and is a subset
of the survey area for the 2dF galaxy redshift survey.

\begin{figure}
\centering \centerline{\psfig{file=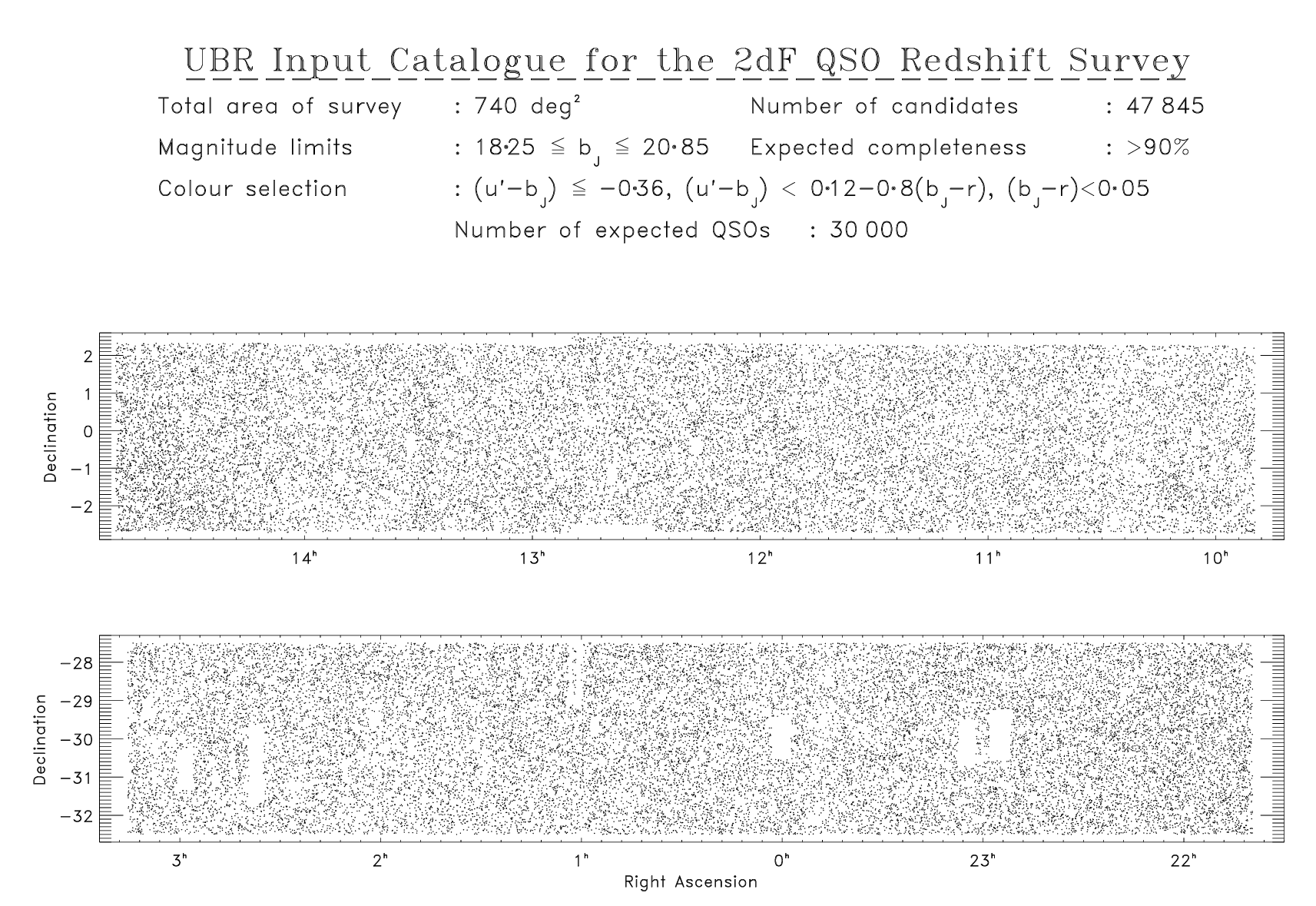,angle=90,width=14cm}}
\caption{The 2dF QSO survey input catalogue based on the selection of
stellar objects from $UB_JR$ UK Schmidt photographic material.  Holes
in the data-set represent areas excised due to bright stars}
\end{figure}

In each UKST field APM measurements of one $J$ plate, one $R$ plate and
up to four $U$ plates/films were used to generate a catalogue of
stellar objects with $B_J < 20.85$.  A sophisticated procedure was
devised to ensure catalogue homogeneity across plate boundaries (Smith
1998).  The criteria for inclusion in the catalogue were $(U-B_J) \le
-0.36$; $(U-B_J)<0.12-0.8(B_J-R)$; $(B_J-R) < 0.05$.  The full
$UB_JR$-selected catalogue is shown in figure 5.  Based on the colours
of QSOs which have previously been identified in the survey region, we
estimate the catalogue is $\sim 93$\% complete for $z<2.2$ QSOs and
comprises over 55\% QSOs (see Croom 1997). Preliminary indications
from the 2dF survey confirm this QSO fraction, with the principal
contamination arising from galactic subdwarfs and compact blue
galaxies.

\subsection{Results} 

\begin{figure}
\centering \centerline{\psfig{file=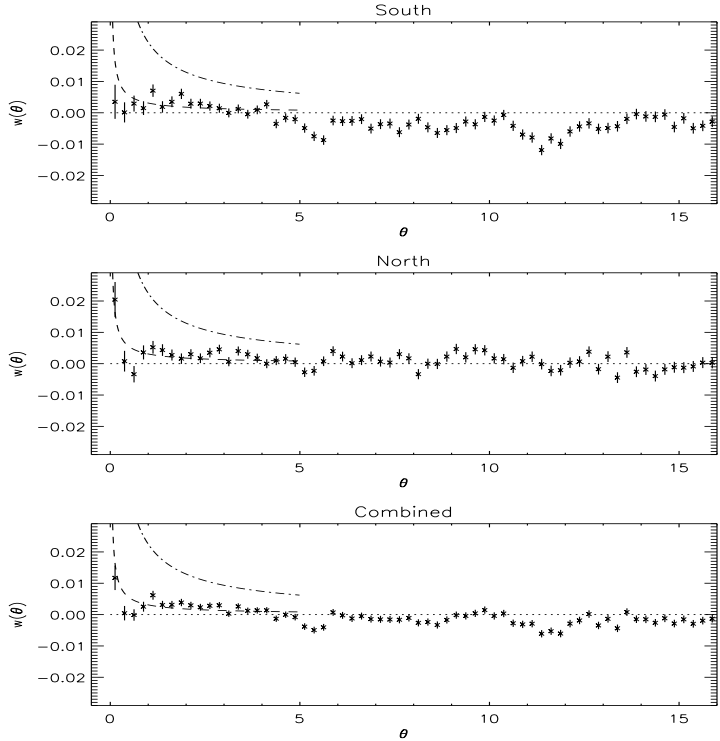,height=10cm,width=10cm}}
\caption{The auto-correlation function for the 2dF QSO survey input
catalogue. The curves denote the expected angular correlation function
assuming $r_0=6$h$^{-1}$Mpc (dashed line) or $r_0=18$h$^{-1}$Mpc
(dot-dashed) for a $-1.8$ power law spatial correlation function 
and stable clustering in comoving space.}
\end{figure}

Smith (1998) has measured the angular auto-correlation function,
$w_{\rm QQ}(\theta)$ from the 2dF QSO input catalogue. To minimise the
contamination from sources other than QSOs, Smith (1998) restricted
the auto-correlation analysis to objects in the input catalogue with
$U-B_J< -0.5$.  Based on the derived contamination rates, this
restricted sample comprises almost 80\% QSOs.

The angular auto-correlation function derived by Smith (1998)
is plotted in figure 6. Smith (1998) reports a 3$\sigma$ detection of 
clustering consistent with a scale length of $r_0 = 4.9 \pm 
1.0\,$h$^{-1}$Mpc, assuming the scale length of the QSO correlation
remains constant in comoving space as a function of redshift.
This is the first detection of QSO clustering on the plane of the
sky from an optically-selected catalogue and the amplitude
is consistent with the scale length of QSO clustering previously
derived from spatial clustering studies.  It is inconsistent
with the much greater scale length of $r_0 \sim 18\,$h$^{-1}$Mpc
inferred for radio-loud QSOs by Yee \& Green (1987). 

At angular scales which are multiples of the mean field-to-field
separation, $5.5^{\circ}$ and $11^{\circ}$, $w_{\rm QQ}(\theta)$ does
show systematic deviations from zero amplitude.  Although these
deviations are small ($<0.005$) they may represent the fundamental
limit of the accuracy to which large-scale structure can be measured
at scales $r \sim 400\,$h$^{-1}$Mpc (the comoving separation
corresponding to $5^{\circ}$ at $z=1.5$).

\section{The 2dF QSO redshift survey}

2dF spectroscopic observations for the QSO redshift survey began
in January 1997, although the bulk of the redshifts have been obtained
in the October 1997 and January 1998 observing runs as the 2dF
system has gained functionality (increased number of fibres, 
faster field re-configuration times).  Over 1000 QSO redshifts have
been obtained with the 2dF, and the 2dF QSO redshift survey is already
the largest single homogeneous QSO catalogue in existence.  Initial 
results from this survey are shown in figures 7--10.  

\begin{figure}
\centering \centerline{\psfig{file=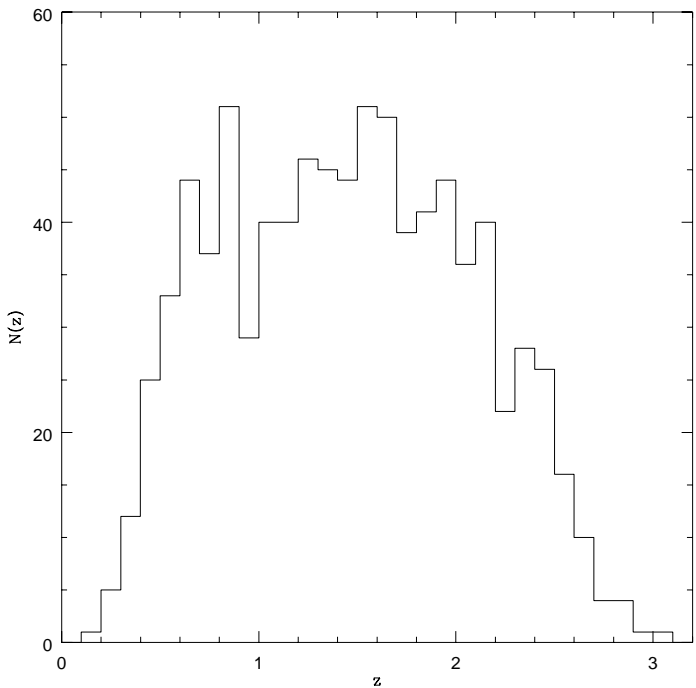,width=10cm}}
\caption{The number-redshift relation for the QSOs identified to 
date (March 1998) in the 2dF QSO redshift survey.}
\end{figure}

\begin{figure}
\centering \centerline{\psfig{file=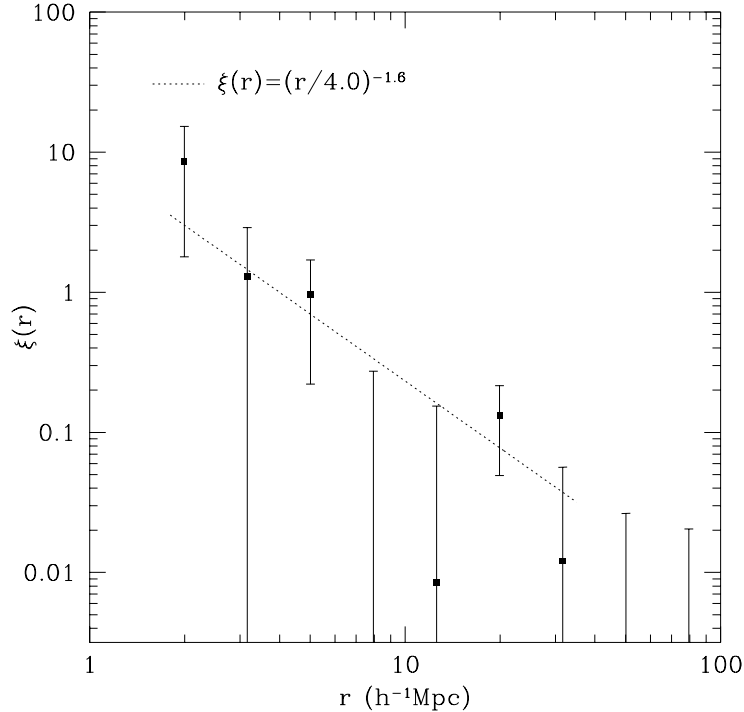,width=10cm}}
\caption{The spatial auto-correlation function for QSO identified
to date in the 2dF QSO redshift survey.  The dotted line denotes 
$\xi_{\rm Q}(r) = (r/4{\rm h}^{-1}{\rm Mpc})^{-1.6}$}
\end{figure}

\begin{figure}
\centering \centerline{\psfig{file=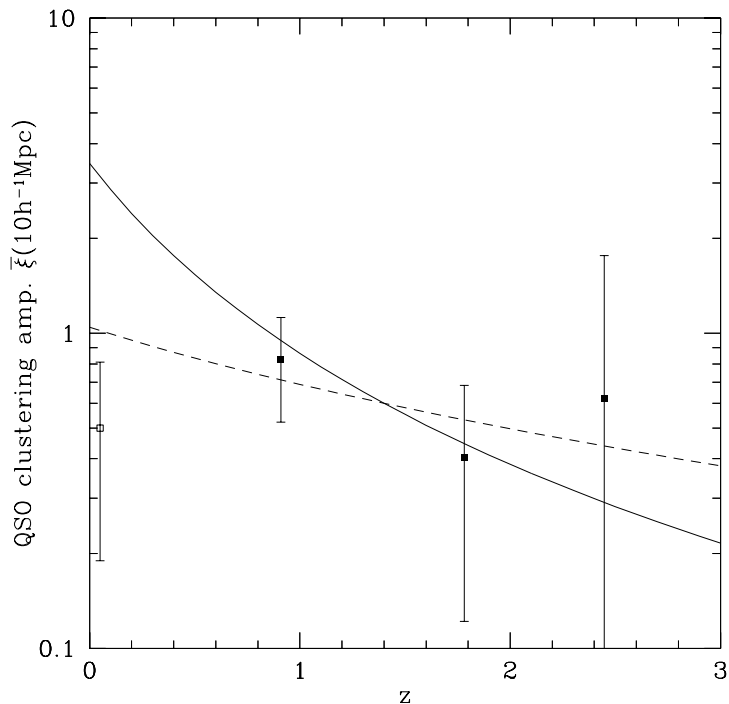,width=10cm}}
\caption{$\bar{\xi}_{\rm Q}(10)$, the mean amplitude of the QSO 
correlation function below $10{\rm h}^{-1}$Mpc, plotted as a 
function of redshift (filled squares).  Estimates are based on the combined  
Durham/AAT, CFHT, LBQS and the 2dF survey data currently available.  
The point at $z\sim0$ (open square) is derived from the AGN
correlation functions of 
Boyle \& Mo (1993) and Georgantopoulos \& Shanks (1994). 
Model predictions for the growth of structure
based on linear theory in an $\Omega_0=1$ (solid line) 
and $\Omega_0=0.1$ Universe (dashed line) are shown. }
\end{figure}

\begin{figure}
\centering \centerline{\psfig{file=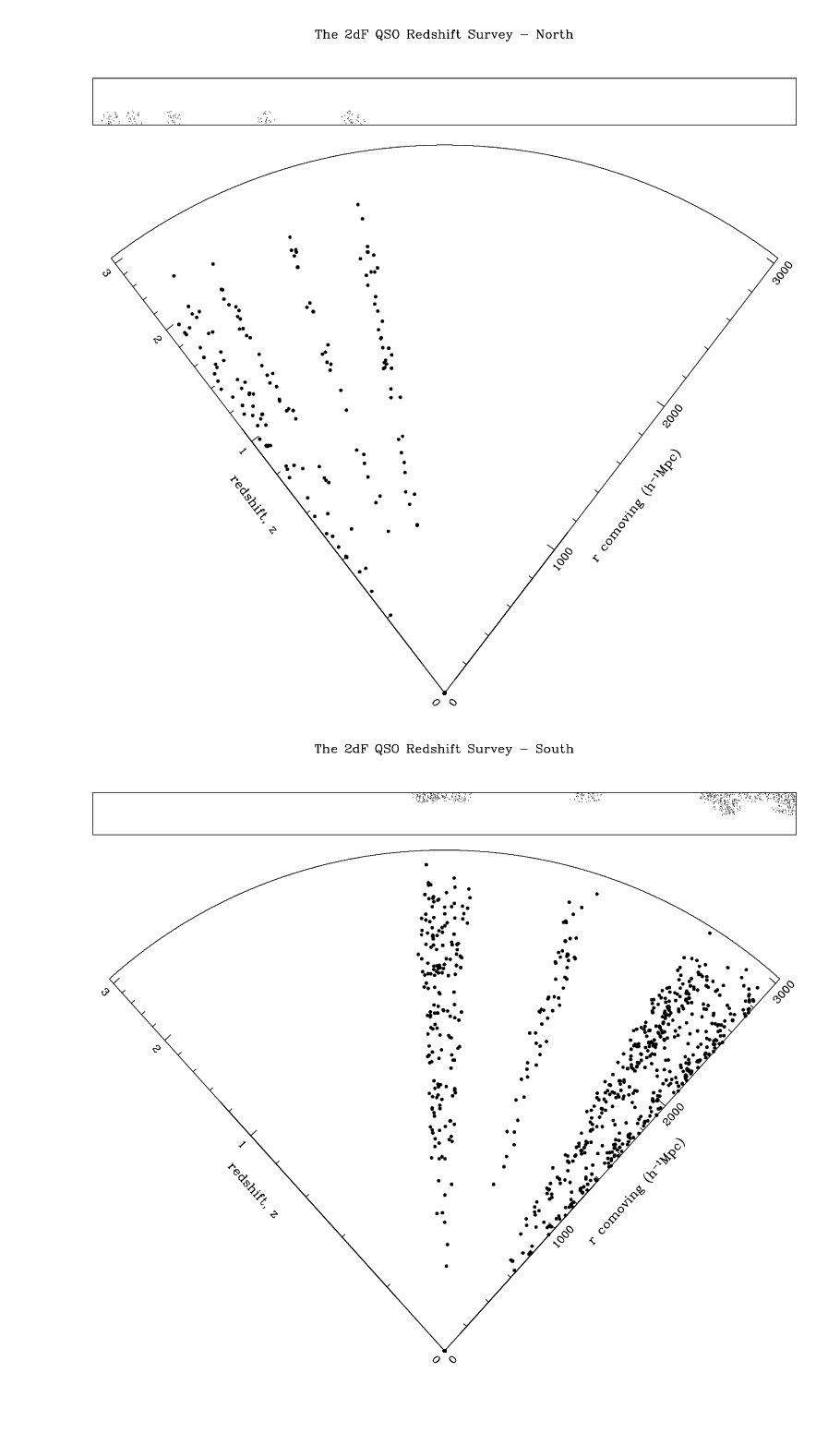,width=12cm}}
\caption{Cone plots for QSOs identified in the $0^{\circ}$ and
$-30^{\circ}$ declination strips from 2dF observations to date.}
\end{figure}

Figure 7 shows the number-redshift, $n(z)$, relation for the sample.
The $n(z)$ shows a broad flat distribution, with no obvious redshift
peaks or troughs due to selection biases.  The auto-correlation
function plotted in figure 8 is consistent with the form previously
derived by Croom \& Shanks (1996); $\xi(r) = (r/4{\rm h}^{-1})^{-1.6}$.
Although the error bars are still large, there is still no strong
evidence for evolution of the correlation length with redshift (figure
9).  The predictions for two linear theory models of structure growth
confirm the result of Croom \& Shanks (1996) that the lack of
evolution favours a low $\Omega_0$ universe in such models.

With 97\% of the 2dF QSO redshift survey remaining to be completed,  
the survey cone diagram plotted in figure 10 is currently rather
empty, but it does serve to illustrate the eventual scale of the
2dF QSO redshift survey.

The input catalogue is also useful as a resource for other QSO studies.
In a separate project, we have cross-correlated the input catalogue
with the NRAO/VLA sky survey (NVSS, Condon {\it et al.} 1998),
yielding over 600 detections (corresponding to 5\% of the QSOs in the
input catalogue).  Keck spectra have been obtained for almost 100 of
these sources, the vast majority of which ($>95$\%) are QSOs.

\begin{figure}
\centering \centerline{\psfig{file=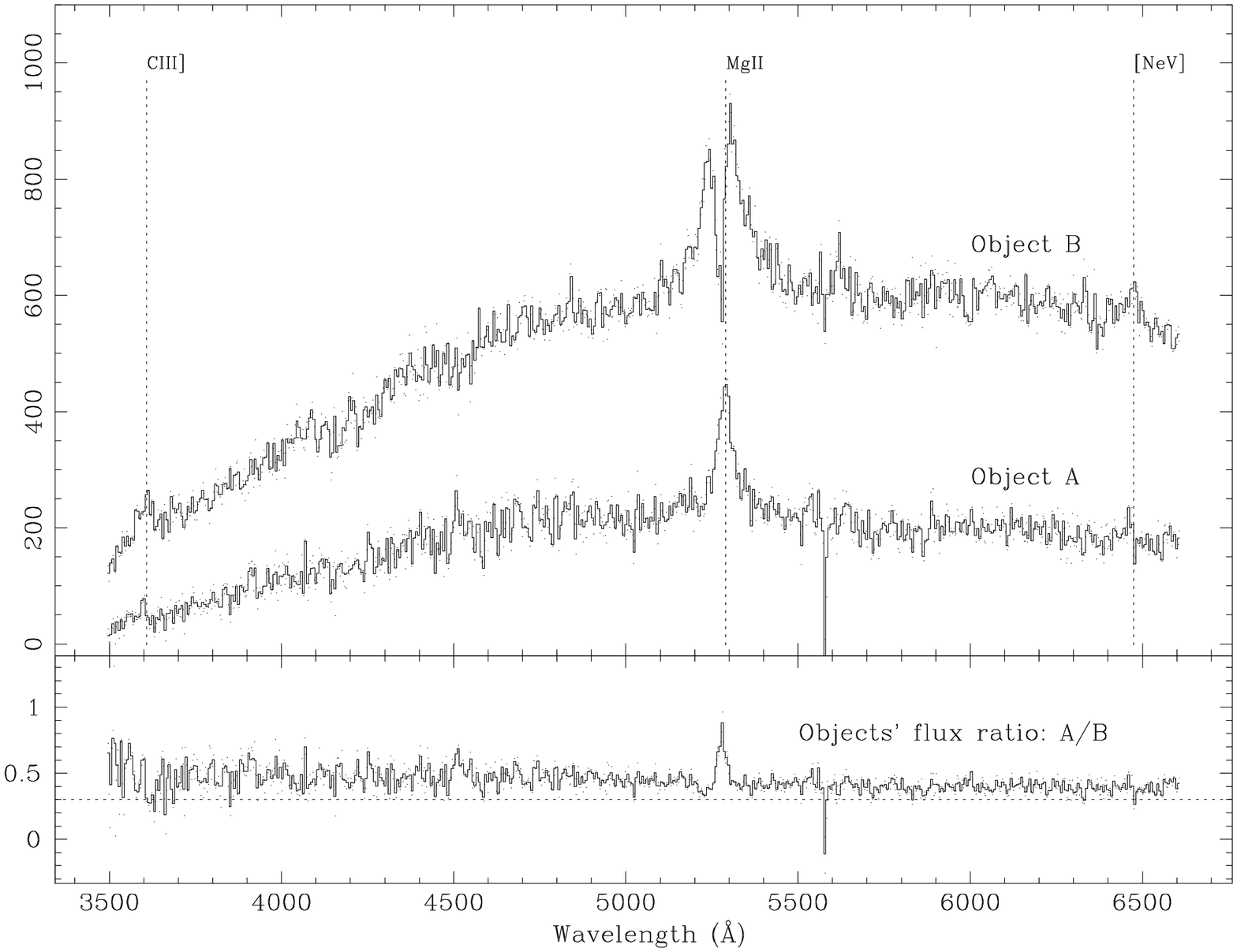,width=14cm,angle=270}}
\caption{Spectra for the QSO pair Q0101.8--3012A,B, with the ratio
spectrum plotted at the bottom.}
\end{figure}

We have also begun a more conventional long-slit spectroscopic program
to obtain redshifts for all input catalogue pairs with angular
separations less than 20$\,$arcsec.  Such objects are candidate
wide-separation gravitational lenses or binary QSOs. The incidence of
wide-separation gravitational lenses can be used to place constraints
on the value of the cosmological constant.  Following 
Maoz {\it et al.} (1997), Croom
(1997) calculates that lensing by `typical' cluster masses ($2 \times
10^{14}$M$_{\odot}$) would result in between four ($\Omega_0=1$,
$\lambda_0=0$) and fourteen ($\Omega_0=0.4$, $\lambda_0=0.6$)
$5-20$-arcsec QSO lenses being seen in the survey.  To date, we have
surveyed approximately 25\% of the catalogue pairs at these
separations, and found only one possible lens candidate (figure 11).
The two components of the possible lens (Q0101.8--3012A,B) have a
separation of $17\,$arcsec on the sky.  At the redshift of the system
($z=0.89$), this corresponds to a projected proper spatial separation
of 71h$^{-1}$kpc.

\section{Future Prospects}
Currently the 2dF QSO redshift survey is only 3\% complete.
Yet it is already the single largest homogeneous QSO redshift
survey yet compiled.  Many of the cosmological tests proposed above
will become possible when the survey is completed, providing
measures of $\Omega_0$, $b_{\rm Q}$ and $\lambda_0$ as well as
new information on the redshift evolution of structure in the
Universe and on LSS at scales of many hundreds to thousands of 
megaparsecs.  In addition the catalogue will provide a 
unique resource for damped Ly$\alpha$ systems, and common absorption
line systems.  It may also prove possible to derive 
information on the power spectrum of mass fluctuations
at $z>2$ through the Ly$\alpha$ forest absorption via the
method of Croft {\it et al.} (1997).

The next decade will see the completion of the 2dF QSO survey together
with the Sloan Digital Sky Survey ($10^5$ QSOs; see Margon, this
volume).  The LAMOST survey ($10^6$ QSOs, see Chu \& Zhao 1997) is
also planned to be well into its operational phase.  When QSOs were
first discovered, the hope was they they could be used to carry out
standard cosmological tests. Now, almost forty years on, with the
advent of these new surveys, this goal may finally be realised.


\smallskip
\end{document}